# Application Layer Intrusion Detection with Combination of Explicit-Rule-Based and Machine Learning Algorithms and Deployment in Cyber-Defence Program

**Amal Saha**[*]
Tata Institute of Fundamental Research (TIFR), Mumbai, INDIA,
Email: amal.k.saha@gmail.com

**Sugata Sanyal**
Tata Consultancy Services (TCS), Mumbai, INDIA
Email: sugata.sanyal@tcs.com

*Corresponding Author

———————————————Abstract———————————————————

There have been numerous works on network intrusion detection and prevention systems, but work on application layer intrusion detection and prevention is rare and not very mature. Intrusion detection and prevention at both network and application layers are important for cyber-security and enterprise system security. Since application layer intrusion is increasing day by day, it is imperative to give adequate attention to it and use state-of-the-art algorithms for effective detection and prevention. This paper talks about current state of application layer intrusion detection and prevention capabilities in commercial and open-source space and provides a path for evolution to more mature state that will address not only enterprise system security, but also national cyber-defence. Scalability and cost-effectiveness were important factors which shaped the proposed solution.

## I. INTRODUCTION

Intrusion detection and prevention at both network and application layers are important for cyber-security and enterprise system security.

There have been numerous works on network intrusion detection and prevention systems, but focus on application layer intrusion detection and prevention has been inadequate and work on this is not very mature.

Since application layer intrusion is increasing day by day, we must address it too and apply state-of-the-art algorithms for effective detection and prevention.

This paper talks about current state of application layer intrusion detection and prevention capabilities in commercial and open-source space and builds a conceptual framework to address not only enterprise system security, but also national cyber-defence. Scalability and cost-effectiveness were taken into consideration to arrive at the conceptual framework.

Future work on this would focus on proof-of-concept implementations by researchers and also more concrete implementations by private enterprises and government agencies.

## II. LITERATURE SURVEY

Application level intrusion detection and prevention system is becoming very important because attack or intrusion at application layer (OSI layer 7) rather than network layer (OSI layer 4) is increasing significantly. Real-time detection and prevention within the process of web application is important. But certain events generated outside the process of web application (e.g., attack on honeypot application in the same site as the web application in question) may also be leveraged to gain further insight through near real-time or offline computation of correlation or other metrics, with the events occurring inside the web application in question. Complexity involved in detecting intrusion increases manifold, when we process events from multiple processes. Even with event data from single process, it may be difficult for many complex use cases. And detection techniques based solely on predefined, explicit rules may be infeasible. Many existing techniques using explicit intrusion detection rules and not using machine learning, have been successful in case of single-process event analysis, e.g., OWASP AppSensor [2, 3], for application layer intrusion detection. The Open Web Application Security Project or OWASP [1] already classified SQL Injection as web application vulnerability and Frank S. Rietta [19] talks about an SQL Injection attack detection technique. Holger Dreger et al [20], in their discussion on dynamic application-layer protocol analysis for network intrusion detection, talked about enhancement in detection of applications not using their standard ports, payload inspection of FTP data transfers, and detection of IRC-based botnet clients and servers, but do not touch upon HTTP protocol level intrusion detection. Trivedi et al [7] talk about reputation based network IDS and conclude that pure IDS alone may not be able to address reputation related to human behaviour in the context of intrusion detection because of complexity.

Intruders employ various techniques to gain access and conceal data and Sandipan Dey et al [18] highlighted a novel technique that intruders could potentially use to hide data or avoid detection.

Automated and scalable learning from large set of events from multiple processes can help compute correlation and other metrics required for intrusion detection and prevention models. Scalable intrusion detection is important in enterprise deployment and also for cyber-security setup of a nation.

Correlation of events from multiple processes is a fundamental log-data analysis use case for unsupervised and reinforcement machine learning. Various machine learning algorithms have been proposed and implemented in existing network intrusion detection and prevention systems. Sampada Chavan et al [5] proposed two machine-learning paradigms, namely, Artificial Neural Networks and Fuzzy Inference System and predicted data correlation as future of intrusion detection system. Zhihua Cui et al [15] described a method of training artificial neural networks. Ajith Abraham et al [8] and alsoChet Langin et al [14] talk about soft computing based IDS techniques and machine learning is considered as subset of soft computing paradigms. Theuns Verwoerd et al [6] highlighted methodologies of statistical models, immune system approaches, protocol verification, file and taint checking, neural networks, whitelisting, expression matching, state transition analysis, dedicated languages, genetic algorithms and burglar alarms. And some of these are related to machine learning and the rest are standard statistical and non-statistical approaches. Unsupervised learning using clustering of data [9] is also used in intrusion detection. S Saravanakumar et al [16] introduce some artificial neural networks algorithms, a type of implementation of machine learning, for network intrusion detection systems. Hung-Jen Liao et al, in their review [17] classify intrusion detection approaches as statistics-based, pattern-based, rule-based and state-based.

Except AppSensor [2, 3], the above mentioned approaches addressed OSI layer-4 attacks so far. Security Incident and Event Management (SIEM) solutions often deployed in enterprise data centers depend on event data generated by multiple applications or processes and necessarily involve multi-step intrusion use cases. SIEM [4] solution acts as real-time monitor of the incident and event data generated by applications and network components and can help in detection and prevention of vulnerabilities and intrusion at network and application levels. SIEM solutions have been widely implemented in enterprises and many new variants of commercial SIEM solutions are using machine learning algorithms. SIEMs are good at detecting multi-step intrusion, but are in general expensive to procure and operate and require significant configuration post deployment for effective detection. SIEM is positioned in the market as a versatile solution that can help detect intrusion at various OSI layers and correlation is considered as one of the defining characteristics of SIEM.

Gartner [10] introduced the concept of runtime application self-protection (RASP) as a way of handling vulnerabilities in web applications. OWASP AppSensor [2, 3] is an open source reference implementation of in-process, application level IDPS. There are commercial implementations of the concept similar to AppSensor, e.g., Application Defender from HP [11], Prevoty Automatic Runtime Application Self-Protection [12]. Major part of AppSensor reference implementations and that of the commercial implementations of RASP are concerned with SQL Injection and Cross-Site-Scripting (XSS) which have been highlighted by Top Ten OWASP [1]. AppSensor [2] talks about many attacks which go beyond OWASP Top Ten vulnerabilities [1]. Intrusion and vulnerabilities at application layer are not the same. Remediation approaches for vulnerabilities of already known patterns have been documented and The Open Web Application Security Project or OWASP [1] has been spearheading this since last 15 years. These vulnerabilities may be remediated by building capabilities within web application in question or delegating the solutions to web application firewall (WAF). WAF sits inline in front of the web application as a separate process and most WAFs can detect the detection vulnerabilities based on signature or known patterns. WAF is often deployed as a reverse proxy. There are many vendors of WAF and WAF has been quite successful in remediation of many OWASP identified vulnerabilities. However, for certain types of vulnerabilities, lack of full context of the HTTP request and the associated session in an out-of-process WAF, makes it somewhat less effective than the one where the remedial measures are built into the web application itself using secure programming and configuration. For example, remediation of privilege escalation is difficult to achieve in an out-of-process WAF because of lack of detailed session information (the context) including the privilege level of the user requesting action on or access to resources on the server. But WAF is a good choice for web applications which went into production without its own remedial measures embedded in application. Except AppSensor [2, 3], the authors feel that other implementations of RASP is focused on OWASP Top Ten [1] and do not qualify to be application layer intrusion detection and prevention systems (IDPSs). This distinction would be elaborated in next paragraphs.

## III. PROPOSED WORK AND COMPARISON WITH OTHER WORKS

Except AppSensor [2,3] which talks about simple algorithms for application layer IDPS features, the authors are not aware of significant discussions that highlight the concept of application layer IDPS, its relevance to the growing field of cyber-defence and use of sophisticated algorithms, machine learning or otherwise, at application level IDPS.

In this paper, we propose IDPS solution approach for application layer intrusion (OSI layer 7) which go beyond SQL Injection and Cross-Site-Scripting (XSS) or more generally, OWASP Top Ten [1].

As the first step, we propose development of a long-term program which will classify and catalogue layer-7 intrusions. Interestingly, open-source AppSensor [2] listed quite a few such intrusion patterns, but it mixed up the intrusion patterns with those covered by WAF and OWASP Top Ten. The proposal here is to cull AppSensor intrusion patterns, after removing OWASP Top Ten vulnerabilities [1] and make living list which would be remediated by own implementations of team of the program owner. The program owners could be government's information security offices which go by various names in different counties, e.g., national security agency, homeland security, cyber-security office, etc.

As the second step, we propose development of a framework that would consist of a module having components that would be part of runtime of the web application in question and would generate events. The events would feed external IDPS system to be built. The external IDPS system would collect events from multiple sources including the web application in question and network IDPS, WAF, etc as optional sources. Instead of buying SIEM from a vendor, this IDPS system would be developed from scratch and must be scalable. Of Course, the development may be outsourced to a specialist vendor with SIEM expertise, but the requirements (the intrusion patterns in step 1) and the design (step 2) would incorporate explicit rule based algorithms and those based on machine learning and the solution should be extensible so that new algorithms may be plugged-in in future. Various patterns or strategies should be identified on a continuous basis for prevention in real-time, near real-time and batch modes and be implemented. Also the solution approach, particularly the in-process runtime components, should be implemented in major programming languages used for development of web applications, e.g., Java Enterprise Edition and Microsoft Dot Net.

As for availability of implementations of machine learning algorithms, Apache Mahout [13] implemented some algorithms in Java and it is typically used for recommendation engine and classification and has so far not been used in intrusion detection. It is expected that the IDPS implementation team would implement the documented machine learning algorithms in the programming language of their choice.

The authors believe that this is how a large-scale and cost-effective solution may be developed to address the question highlighted in this paper.

As for preparation of the list of patterns (step 1), exploitations of the OWASP classified vulnerabilities [1] qualify as attacks per se, on web application. While these vulnerabilities contribute to the risks the web application may be exposed to, attacks which do not necessarily exploit the above classified vulnerabilities are not uncommon. Let us have some examples of attacks which go beyond the known and classified vulnerabilities [2]. High number of logouts or logins in a web application or all web applications in the site, very high or very low speed of use of web application or all web applications in the site, frequency of use of web application by a specific user or device, unusual geo-location or time from which or at which web application is accessed by a given user or device, are some examples of many possible attacks at layer 7 [2] which a standard WAF may not be detecting.

As for preventive action after detection of such attacks or intrusions, there would be various strategies [3] to prevent them by way of response, namely, disabling the requested function, disabling or locking out user account, logging or notifying concerned team to individual, etc. Again AppSensor [3] identified quite a few prevention approaches against attacks which are addressed outside the proposed application level IDPS (i.e., they are standard OWASP Top Ten vulnerabilities [1]). Hence culling would be needed to pick up the first list from AppSensor project [2] and this must be expanded on a continuous basis.

Application level intrusion detection and prevention (i.e., response) can happen, in-process (in the same process hosting the web application) for a large number of use cases (e.g., very high or very low speed of use of web application, or high number of logouts or logins in a specific web application). Advantage of this in-process detection is availability of full request context in the web application runtime. But some use cases (e.g., very high or very low speed of use of web applications in a site across web applications, or high frequency of use of web applications in a site by a group of users at a point in time) may be better handled out-of-process. This can happen if adequate contextual information is given by the web application in question to the detecting and preventing system running outside the individual web applications and the detecting subsystem is capable of performing aggregate function like correlation with multiple parameters. Prevention strategy may involve waiting in the web application to be protected till the out-of-process detection and prevention system sends result of detection and recommendation for prevention. In this case, it is still inline processing. Other out-of-process prevention strategy could be sending recommendation to concerned web applications or network components to block request from a set of users or IP address, in near real time, but not inline. Both explicit rule based and machine learning based techniques for application layer detection of intrusion may be applied in-process and out-of-process configurations mentioned above. Given that Java and Dot Net are two major programming languages for web applications, in-process application of machine learning would require implementation of the selected algorithms in these two major programming languages.

Various governments around the world came up with plan for national critical information infrastructure protection and cyber security is getting significant funding and various governments are coming up with national approach for tackling the issues. It is because often the adversaries are nation state and agencies sponsored by other governments with high funding capability. Intrusion detection is a critical component of homeland security [21]. Besides firewall and network intrusion detection (layer 3 and 4) systems, it is important to detect layer 7 intrusions and find approaches for remediation. Since national interests are at stake, authors think that generic solution approaches highlighted here may be implemented by entities handling national critical information infrastructure. Authors suggest implementation of similar approaches with enhancement through development of new algorithms based on combination of signature (explicit rules for detection) and machine learning. The application in the proposed system acts as one of the sensors collecting intrusion data and feeding the detecting and preventing components.

Many network based IDPS suffer from the fact that they often get encrypted packet when HTTPS is enabled between the browser or HTTP client application and the server configured to serve HTTPS requests. Network IDPS under this configuration is blind to encrypted packets. If IDPS capabilities are built in web application or event data for IDPS is served by the web application, the limitation in the process of detection of intrusion goes away. Hence remediation of standard OWASP vulnerabilities by incorporating OWASP prescribed remedies in web application itself or through delegation to WAF and RASP, is the first step towards layer 7 defences against intrusion. The next layer is provided by building proposed application level intrusion detection and prevention approaches discussed in this paper.

## IV. ARCHITECTURAL COMPONENTS AND DESIGN

**Components of the proposed system are the following:**

- A. Firewall ( OSI layer 4)

- B. Network Intrusion Detection and Prevention System (Network IDPS)

- C. WAF - Web Application Firewall (layer 7)

D. Runtime IDPS Components (App Layer) - standard and machine learning components for in-app detection, prevention plus event generation for external IDPS system, with supporting messaging components for interaction with external IDPS system

E. IDPS System - with standard and machine learning components and may have bidirectional interactions with WAF and web application with runtime IDPS capabilities

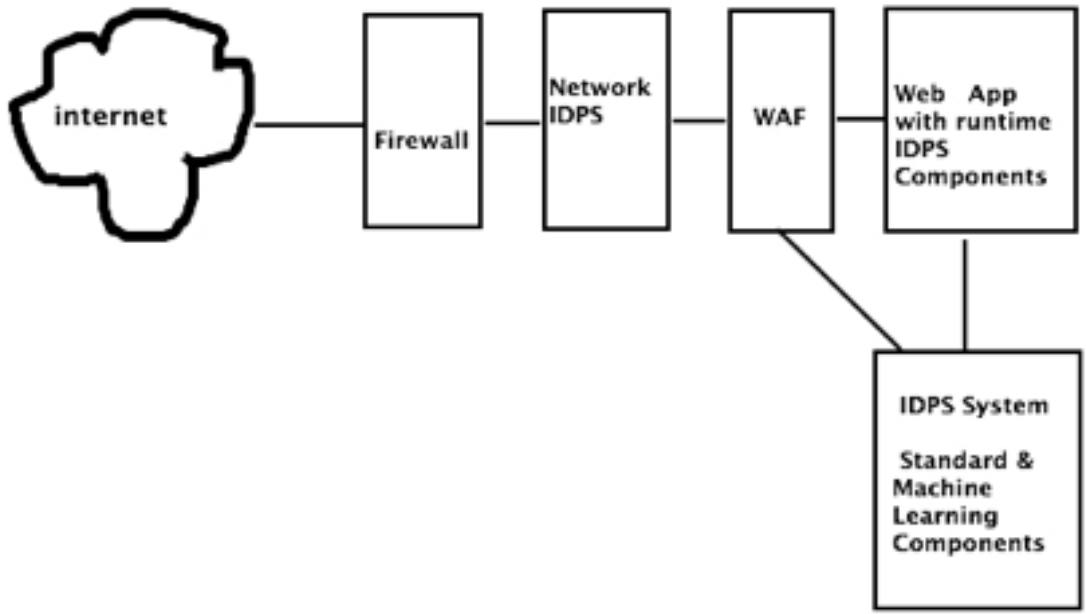

Figure 1 : Architecture of Proposed System

## V. ASSUMPTIONS AND CONCEPTUAL STEPS

Implementation of machine learning (ML) and standard (non-ML) algorithms in proof-of-concept implementation requires significant computation and would be the subject matter of future papers by the authors. In this paper, we are limiting ourselves to related assumptions, high level steps in proposed algorithms and preliminary analysis.

**Assumptions**
I. Application layer intrusion use cases are identified to a large extent (cannot however be assumed to be exhaustive) and understood before developing algorithm.
II. Adequacy of Machine Learning (ML) and non-ML (standard) algorithms would be tested on the list of use cases for training over a long period of time in laboratory environment and would be refined progressively, through testing.

**Conceptual Steps**
I. HTTP(S) request data arrives at the server and data about user access already exists in the database of the web application. Let us take a use case - frequency of web-site use by a given user. System learnt the usage pattern in course of time using standard algorithm SA1. Since it is a relatively simple use case, it need not depend on machine learning and standard explicit coding of rule may work. Usage pattern data is stored in database of the web application, in a schema or representation demanded by the algorithm.

II. The algorithm is positioned as a filter in the chain of processing that takes place to service the request.
III. The algorithm analyses the current access request with respect to historical access pattern data, after retrieving the data from database or application cache.
IV. If the algorithm should correlate the data with usage data from firewall or network intrusion detection system, incorporate that in the algorithm. Make the algorithm flexible, by introducing parameters to control the behaviour.
V. Algorithm already defined allowed variance of usage and it would use that metric to determine if the current request frequency sounds abnormal or not.
VI. Review the result with expectation or test case already prepared before testing the algorithm, for the intrusion use case.
VII. Repeat the procedure with machine learning algorithm, MLA1, implemented by the team and see if this is more effective, statistically speaking, by determining success rate against test cases for the given use case. If MLA1 is better than SA1, select it for use in production deployment. Vipin Das et al [22] applied machine learning algorithms Rough Set Theory (RST) and Support Vector Machine (SVM) to detect network intrusions where RST is used to pre-process the data and reduce the dimensions, and then the features selected by RST is sent to SVM model to learn, and they reported some success in detection. They [22] also referred to Principal Component Analysis algorithm. Various such ML algorithms may be shortlisted and attempted for application layer IDPS proof-of-concept to cover the use cases listed in step one (section III).
VIII. There could be use case where non-ML algorithm would not be suitable at all and in that case start with ML algorithm only. Certain ML algorithms would require supervised learning and would require intervention of an expert in data preparation.

## VI. CONCLUSION

This paper talks about the design of a state-of-the-art, conceptual framework for building application layer intrusion detection and prevention and the solution is supposed to be scalable and cost-effective. The implementation of the framework requires significant computation and future work would focus on building proof-of-concept implementations by researchers to validate the conceptual framework discussed here. Implementation of machine learning algorithms and adoption is growing in mainstream programming languages and this would contribute to concrete implementation of the framework.

Since application layer intrusion is increasing day by day, importance of application layer intrusion detection would grow significantly. Various organizations, private enterprises and government agencies, would take up concrete implementation of the conceptual framework discussed here.

## REFERENCES

[1] Open Source forum for security projects related to web applications - OWASP https://www.owasp.org/ and Top Ten classified vulnerabilities https://www.owasp.org/index.php/Top_10_2013-Top_10

[2] Michael Coates, Dennis Groves, John Melton, Colin Watson, OWASP AppSensor Detection Points - https://www.owasp.org/index.php/AppSensor_DetectionPoints

[3] Michael Coates, Dennis Groves, John Melton, Colin Watson, OWASP AppSensor Detection Points, OWASP AppSensor Response Actions - https://www.owasp.org/index.php/AppSensor_ResponseActions

[4] Alex Pinto, Secure because Math: A deep-dive on Machine Learning-based Monitoring, Black Hat Briefings USA 2014.

[5] Sampada Chavan, Khusbu Shah, Neha Dave, Sanghamitra Mukherjee, Ajith Abraham, Sugata Sanyal, Adaptive Neuro-Fuzzy Intrusion Detection Systems, Proceedings of the International Conference on Information Technology: Coding and Computing, ITCC 2004, volume 1, IEEE

[6] Theuns Verwoerd and Ray Hunt, Intrusion Detection Techniques and Approaches, Computer Communications, Volume 25, Issue 15, 15 September 2002, Pages 1356–1365.

[7] Animesh Kr Trivedi, Rishi Kapoor, Rajan Arora, Sudip Sanyal and Sugata Sanyal, RISM - Reputation Based Intrusion Detection System for Mobile Adhoc Networks, 3rd International Conference on Computers and Devices for Communication (CODEC-06) CNA Institute of Radio Physics and Electronics, University of Calcutta, December 18-20, 2006

[8] Ajith Abraham, Ravi Jain, Sugata Sanyal, Sang Yong Han, SCIDS: A Soft Computing Intrusion Detection System, book "Distributed Computing-IWDC 2004", pages 252-257, Springer Berlin Heidelberg

[9] Ravi Ranjan and G. Sahoo, A New Clustering Approach for Anomaly Intrusion Detection, International Journal of Data Mining & Knowledge Management Process (IJDKP) Vol.4, No.2, March 2014

[10] Joseph Feiman, Neil MacDonald, Gartner's Magic Quadrant for Application Security Testing, published 1st July, 2014

[11] Application Defender from Hewlett-Packard, http://www.gartner.com/technology/media-products/pdfindex.jsp?g=hp_security

[12] Automatic Runtime Application Self-Protection from Prevoty, https://www.prevoty.com/product/overview

[13] Machine Learning algorithms, reference implementations in programming language Java, Apache Mahout, https://mahout.apache.org

[14] Chet Langin · Shahram Rahimi, Soft computing in intrusion detection: the state of the art, J Ambient Intell Human Comput (2010) 1:133–145

[15] Cui, Zhihua, Chunxia Yang, and Sugata Sanyal. "Training artificial neural networks using APPM." International Journal of wireless and mobile computing 5.2 (2012): 168-174.

[16] S. Saravanakumar, Umamaheshwari, D. Jayalakshmiand R. Sugumar, Development and Implementation of Artificial Neural Networks for Intrusion Detection in Computer Network, International Journal of Computer Science and Network Security, VOL.10 No.7, July 2010

[17] Hung-Jen Liao, Chun-Hung Richard Lin, Ying-Chih Lin, Kuang-Yuan Tung, Intrusion detection system: A comprehensive review, Journal of Network and Computer Applications 36 (2013) 16–24

[18] Sandipan Dey, Ajith Abraham, and Sugata Sanyal. "An LSB Data Hiding Technique Using Natural Number Decomposition." Intelligent Information Hiding and Multimedia Signal Processing, 2007. IEEE Third International Conference on Intelligent Information Hiding and Multimedia Signal Processing, IIHMSP 2007, Nov 26-28, 2007, Kaohsiung City, Taiwan, IEEE Computer Society press, USA, ISBN 0-7695-2994-1, pp. 473-476, 2007.

[19] Frank S. Rietta, Proceedings of the 44th annual South-east regional conference, 2006, Pages 531-536, publisher ACM

[20] Holger Dreger, Anja Feldmann, Michael Mai, Vern Paxson, Robin Sommer, Dynamic Application-Layer Protocol Analysis for Network Intrusion Detection, 15th USENIX Security Symposium, Pages CHANGE 257–272 of the Proceedings, 2006

[21] Abraham, Ajith, and Johnson Thomas. "Distributed intrusion detection systems: a computational intelligence approach." Applications of information systems to homeland security and defense. USA: Idea Group Inc. Publishers (2005): 105-35

[22] Vipin Das, Vijaya Pathak, Sattvik Sharma, Sreevathsan, MVVNS.Srikanth, Gireesh Kumar T, Amrita Vishwa Vidyapeetham, Network Intrusion Detection System Based on Machine Learning Algorithms, International Journal of Computer Science & Information Technology (IJCSIT), Vol 2, No 6, December 2010